\documentclass[pra,twocolumn,amsmath,amssymb,floatfix,groupaddresses,superscriptaddress]{revtex4}
\usepackage{graphicx}
\usepackage{latexsym}
\usepackage{verbatim,times,bbm}
\usepackage{color}

\begin{document}

\title{Entanglement frustration in multimode Gaussian states}

\author{Cosmo Lupo}
\affiliation{School of Science and Technology, Universit\`a di Camerino, I-62032 Camerino, Italy}

\author{Stefano Mancini}
\affiliation{School of Science and Technology, Universit\`a di Camerino, I-62032 Camerino, Italy}
\affiliation{INFN, Sezione di Perugia, I-06123 Perugia, Italy}

\author{Paolo Facchi}
\affiliation{Dipartimento di Matematica and MECENAS, Universit\`a di Bari, I-70125  Bari, Italy}
\affiliation{INFN, Sezione di Bari, I-70126 Bari, Italy}

\author{Giuseppe Florio}
\affiliation{Dipartimento di Fisica and MECENAS, Universit\`a di Bari, I-70126  Bari, Italy}
\affiliation{INFN, Sezione di Bari, I-70126 Bari, Italy}

\author{Saverio Pascazio}
\affiliation{Dipartimento di Fisica and MECENAS, Universit\`a di Bari, I-70126  Bari, Italy}
\affiliation{INFN, Sezione di Bari, I-70126 Bari, Italy}

\begin{abstract}
Bipartite entanglement between two parties of a composite quantum system 
can be quantified in terms of the purity of one party and there always 
exists a pure state of the total system that maximizes it (and minimizes 
purity). When many different bipartitions are considered, the requirement 
that purity be minimal for all bipartitions gives rise to the phenomenon 
of \emph{entanglement frustration}. This feature,  observed in quantum 
systems with both discrete and continuous variables, can be studied by 
means of a suitable cost function whose minimizers are the {\it maximally 
multipartite-entangled states} (MMES). In this paper we extend the analysis 
of multipartite entanglement frustration of Gaussian states in multimode 
bosonic systems. We derive bounds on the frustration, under the constraint 
of finite mean energy, in the low and high energy limit.
\end{abstract}

\maketitle


\section{Introduction}\label{intro}

The development of quantum information theory and technologies has
stimulated and motivated the scientific efforts toward the full
characterization of the geometry of the set of states of quantum
systems \cite{qmath,marmo1,marmo2,marmo3}. For both fundamental and technological reasons,
the interest has been focused on the characterization of entangled
states, that is, those states of a composite system exhibiting
nonclassical correlations among their parties. While the case of
bipartite systems has been extensively studied and --- at least for
pure states --- is nowadays well understood \cite{bient}, the case
of genuinely multipartite entanglement is still not fully mastered. 

The present contribution deals with the problem of multipartite-entanglement 
characterization by focusing on the phenomenon of \emph{entanglement frustration} \cite{frustnjp}. 
In particular, we will explore Gaussian states of continuous variable (CV) systems, 
that is, systems of (quasi-free) quantum harmonic oscillators \cite{CV}. 
Specifically, we consider a suitable cost function, the {\it potential of multipartite entanglement}, introduced in \cite{MMES1}
and extended to the Gaussian framework in \cite{noi}, as a quantifier of frustration.
The fact that this cost function cannot saturate its minimum value is a 
symptom of a sort of frustration of entanglement, induced by
the geometry of the quantum phase space, which prevents the states to be
maximally bipartite-entangled among all the possible system bipartitions.
Here we consider a family of entanglement cost-functions, generalizing
the one introduced in \cite{noi}.
We hence derive new results on their minima in the low and high 
energy limits.

\section{Gaussian entanglement}\label{Gaussian}

Our analysis focuses on a system of $n$ quantum
harmonic oscillators, namely a set of $n$ bosonic modes, described
by the canonical variables $\mathbf{X}=(X_1, X_2, \dots X_{2n}):=(q_1,
q_2, \dots q_n, p_1, p_2, \dots p_n)$. For the sake of simplicity, we
assume all oscillators to be identical, although distinguishable, with
unit frequency, and set $\hbar = 1$. 
We follow \cite{noi} and consider the manifold of Gaussian states of the 
$n$-mode system. 
Let us recall that a Gaussian state $\rho$ is characterized by the first and 
second moments of the canonical variables, that is,
the mean $\langle \mathbf{X} \rangle := \mathrm{tr} (\rho \mathbf{X})$, and the covariance matrix (CM)
$\mathbb{V}$, with elements 
$\mathbb{V}_{a,b} = \frac{1}{2} \langle X_a X_b + X_b X_a \rangle$.
We assume, without loss of generality, $\langle \mathbf{X} \rangle =0$,
and restrict our attention to pure states. The CM of a pure state can be written in the form
\begin{equation}
\mathbb{V} = \frac{1}{2} \mathbb{R} \mathbb{T}^2 \mathbb{R}^\mathsf{T} \, ,
\end{equation}
with
\begin{equation}\label{OS}
\mathbb{T} = \left(\begin{array}{cc} \mathbb{K} & \mathbb{O} \\
\mathbb{O} & \mathbb{K}^{-1} \end{array}\right) \, ,
\qquad
\mathbb{R} = \left(\begin{array}{cc} \mathbb{X} & \mathbb{Y} \\
-\mathbb{Y} & \mathbb{X} \end{array}\right) \, ,
\end{equation}
where $\mathbb{K}$ is diagonal and nonsingular, $\mathbb{O}$ denotes the
null matrix, and $\mathbb{R}$ is a symplectic orthogonal matrix, characterized by the property
that $\mathbb{U}=\mathbb{X}+i\mathbb{Y}$ is unitary.
Finally, we impose a bound on the mean energy per mode, that is,
\begin{equation}\label{energy}
\frac{\mathbb{V}_{k,k} + \mathbb{V}_{n+k,n+k}}{2} \leqslant N +
\frac{1}{2},\quad \forall\; k=1,\dots n
\end{equation}
with $N$ the number of mean excitations per mode. 
This is one of a number of physical constraints that must be imposed on the system in
order to make the problem mathematically (and physically) well posed.
A different approach to entanglement frustration, that makes no use of energy constraints, has been proposed in \cite{adesso}.

We consider {\it purity} as an estimator of bipartite entanglement \cite{MMES1,noi}.
Given a bipartition into two subsystems $\{ A, \bar{A} \}$, associated 
with two subsets of respectively $|A|=n_A$ and $|\bar{A}|=n_{\bar{A}}$ 
bosonic modes ($n_A+n_{\bar{A}}=n$), the purity of the subsystems reads
\begin{equation}\label{purity}
\pi_A = \pi_{\bar A} = \frac{(1/2)^{n_A}}{\sqrt{\det{\mathbb{V}_A}}} \, ,
\end{equation}
where $\mathbb{V}_A$ is the sub-matrix of the CM identified
by the indices belonging to subsystem $A$. Here we have assumed
without loss of generality $n_A \leqslant n_{\bar{A}}$.
It can be easily proven that under the constraint (\ref{energy})
the minimum value of the purity is \cite{noi}
\begin{equation}\label{demoninator}
\pi^\mathrm{min}_{n_A}(N) = \frac{(1/2)^{n_A}}{(N+1/2)^{n_A}} \, .
\end{equation}
The range of the purity is $\pi_A \in [\pi^\mathrm{min}_{n_A}(N),1]$, 
where the value $1$ characterizes {\it factorized states}, 
and the value $\pi^\mathrm{min}_{n_A}(N)$ characterizes those states 
which are {\it maximally bipartite-entangled} across the bipartition considered.

\subsection{Entanglement frustration in multimode systems}

In order to study  multipartite entanglement we introduce the {\it normalized potential of multipartite entanglement}, 
a cost function defined for any pair $(n,n_A)$ by the (normalized) expectation
value $\mathbb{E}$ of the purity over all possible bipartition of given size:
\begin{align}
\chi_{(n,n_A)}(N) & := \frac{1}{\pi^\mathrm{min}_{n_A}(N)} \mathbb{E}\left[ \pi_A \right] \nonumber \\
& = {n \choose n_A}^{-1} \sum_{|A|=n_A} \frac{\pi_A}{\pi^\mathrm{min}_{n_A}(N)} \nonumber \\
& = (N+1/2)^{n_A} \mathbb{E}\left[\det(\mathbb{V}_A)^{-1/2}\right] \, .
\end{align}
The range of the cost function is contained in the interval $[1,1/\pi^\mathrm{min}_{n_A}(N)]$,
where the lower bound characterizes the so-called {\it perfect MMES} 
(Maximally Multipartite-Entangled States) \cite{noi}, which are maximally bipartite-entangled
across all bipartitions of size $n_A$.
However, as shown in \cite{noi}, the geometry of the manifold of CV Gaussian states 
prevents the minimum of the cost function to saturate the  lower bound,
that is, 
\begin{equation}
\chi^\mathrm{min}_{(n,n_A)}(N):= \min  \chi_{(n,n_A)}(N)  > 1 ,
\label{frustrb}
\end{equation}
a feature that is interpreted as 
{\it frustration of entanglement}.

\begin{figure}
\centering
\includegraphics[width=0.4\textwidth]{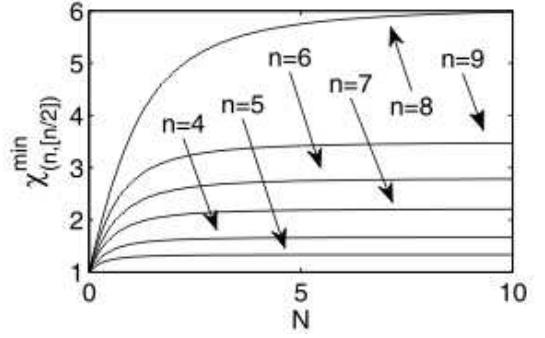}
\caption{Minimum of the normalized potential of multipartite entanglement $\chi^\mathrm{min}_{(n,[n/2])}$ 
vs number of excitations per mode $N$, for $n=4,5,6,7,8,9$.
In the region $N \ll 1$ the minimum value is linear in $N$, while it reaches a plateau for $N \gg 1$.
See Sec.\ \ref{bounds} for details.\label{chi_Nleft}}
\end{figure}

\begin{figure}
\centering
\includegraphics[width=0.42\textwidth]{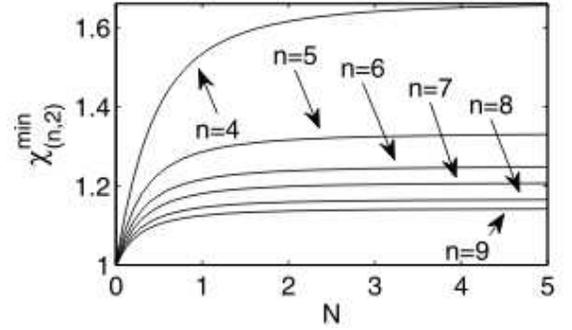}
\caption{Minimum of the normalized potential of multipartite entanglement 
in the case of unbalanced bipartitions $\chi^\mathrm{min}_{(n,2)}$ versus $N$, for $n=4,5,6,7,8,9$.
In the region $N \ll 1$ the minimum value is linear in $N$, while it reaches a plateau for $N \gg 1$.
See Sec.\ \ref{bounds} for details.\label{chi_Nright}}
\end{figure}

Numerical results for the case of balanced bipartitions ($n_A=[n/2]$) were 
presented in \cite{noi}.
Figures \ref{chi_Nleft}, \ref{chi_Nright} shows the behavior of $\chi^\mathrm{min}_{(n,n_A)}$ as a function of $N$ for $n_A=[n/2]$, and for $n_A=2<[n/2]$.
$\chi^\mathrm{min}_{(n,n_A)}(N)$ appears to be a monotonic function of  $N$. 
Two limiting regimes are identified, corresponding to $N \ll 1$ and $N \gg 1$: 
in the former case we notice a linear regime for increasing values of $N$; in the latter, 
$\chi^\mathrm{min}_{(n,n_A)}$ saturates to a constant that depends on the value of $n$.
Moreover, $\chi^\mathrm{min}_{(n,n_A)}(N)$ appears to be a decreasing function of $n$ 
for a given $n_A<[n/2]$, while for $n_A = [n/2]$ it increases with $n$, although oscillating 
between even and odd $n$. 

More generally, frustration --- which is naturally quantified by $\chi^\mathrm{min}_{(n,n_A)}(N)$ --- 
decreases with $n$ at fixed $n_A$, and increases with $n$ if $n_A$ scales 
linearly with $n$ (e.g.\, for balanced bipartitions).
This behavior of frustration in CV Gaussian states is analogous to that observed in 
discrete-variable quantum systems \cite{frustnjp}.

\section{Bounds on entanglement frustration in Gaussian states}\label{bounds}

In order to estimate the bounds on $\chi^\mathrm{min}_{(n,n_A)}(N)$, we restrict our attention to
states such that $\mathbb{K} = e^r \mathbb{I}_n$ (same squeezing for all modes), whose CM reads
\begin{eqnarray}
\mathbb{V} = \frac{e^{2r}}{2} \left(\begin{array}{cc}
\mathbb{X}\mathbb{X}^\mathsf{T} & -\mathbb{X}\mathbb{Y}^\mathsf{T}
\\ - \mathbb{Y}\mathbb{X}^\mathsf{T} &
\mathbb{Y}\mathbb{Y}^\mathsf{T}\end{array}\right) +
\frac{e^{-2r}}{2}\left(\begin{array}{cc}
\mathbb{Y}\mathbb{Y}^\mathsf{T} & \mathbb{Y}\mathbb{X}^\mathsf{T} \\
\mathbb{X}\mathbb{Y}^\mathsf{T} & \mathbb{X}\mathbb{X}^\mathsf{T}
\end{array}\right)\, , \label{class}
\end{eqnarray}
or, equivalently,
\begin{align}
\mathbb{V} = & \frac{\cosh{(2r)}}{2} \mathbb{I}_{2n} \nonumber \\
& + \frac{\sinh{(2r)}}{2}\left(\begin{array}{cc}
\mathbb{X}\mathbb{X}^\mathsf{T}-\mathbb{Y}\mathbb{Y}^\mathsf{T} & -\mathbb{X}\mathbb{Y}^\mathsf{T}-\mathbb{Y}\mathbb{X}^\mathsf{T} \\
-\mathbb{Y}\mathbb{X}^\mathsf{T}-\mathbb{X}\mathbb{Y}^\mathsf{T} &
\mathbb{Y}\mathbb{Y}^\mathsf{T}-\mathbb{X}\mathbb{X}^\mathsf{T}
\end{array}\right) \, . \label{class_}
\end{align}
The energy constraints read 
\begin{equation}
\cosh{(2r)} \le 2N +1 \, .
\end{equation}
In the following we denote respectively $\tilde\chi_{(n,n_A)}$ and $\tilde\chi^\mathrm{min}_{(n,n_A)}$ 
the normalized potential of multipartite entanglement and its minimum evaluated for this family of states.

\begin{table}[ht]
\caption{Numerical estimates of the slopes $\alpha_{(n,n_A)}$ in Eq.\ (\ref{eq:alpha}), 
$\tilde\alpha_{(n,n_A)}$ in Eq. (\ref{eq:alphatilde}), of the minimum of the normalized 
potential of multipartite entanglement for $N=10$, $\chi_{(n,[n/2])}^\mathrm{min}(N=10)$ [which approximates $\lim_{N\to\infty} \chi_{(n,[n/2])}^\mathrm{min}(N)$], and of 
the upper bound $\beta_{(n,n_A)}$ in Eq.\ (\ref{boundinfty}).
The results are obtained for the case of balanced bipartitions, $n_A=[n/2]$.
\label{tabN}}
{\begin{tabular}{@{}ccccc@{}} \toprule
$n$ & $\alpha_{(n,[n/2])}$ & $\tilde\alpha_{(n,[n/2])}$ & $\chi_{(n,[n/2])}^\mathrm{min}(N=10)$ & $\beta_{(n,[n/2])}$ \\ \colrule
4   & 1.33        		 & 1.333333                  & 1.663650                              & 1.666667                     \\
5   & 1.00   	           & 1.000000                   & 1.332326                              & 1.333333                     \\
6   & 2.40   	           & 2.400000                   & 2.780918                              & 2.795085                     \\
7   & 2.00  	           & 2.000000                   & 2.203228                              & 2.213586                     \\
8   & 3.43  	           & 3.428571                   & 5.980689                              & 6.074700                     \\
9   & 3.00  	           & 3.000000                   & 3.470522                              & 3.491497                     \\ \botrule
\end{tabular}}
\end{table}

\subsection{The linear regime, $N \ll 1$}

From Eq.~(\ref{class_}) we get
\begin{align}
\tilde\chi_{(n,n_A)} & =  \nonumber \\
& \hspace{-0.5cm} (2N+1)^{n_A} \mathbb{E}\left[\det\left(\cosh{(2r)}\mathbb{I}_{n_A}+\sinh{(2r)}\mathbb{Z}_A\right)^{-1/2}\right] \, , \nonumber
\end{align}
where we defined
\begin{eqnarray}
\mathbb{Z} := \left(\begin{array}{cc}
\mathbb{X}\mathbb{X}^\mathsf{T}-\mathbb{Y}\mathbb{Y}^\mathsf{T} & -\mathbb{X}\mathbb{Y}^\mathsf{T}-\mathbb{Y}\mathbb{X}^\mathsf{T} \\
-\mathbb{Y}\mathbb{X}^\mathsf{T}-\mathbb{X}\mathbb{Y}^\mathsf{T} &
\mathbb{Y}\mathbb{Y}^\mathsf{T}-\mathbb{X}\mathbb{X}^\mathsf{T}
\end{array}\right) \, ,
\end{eqnarray}
and $\mathbb{Z}_{A}$ denotes the sub-matrix corresponding to the subset
of modes $A$.
Then,
\begin{align}
\tilde\chi_{(n,n_A)} & = \left(\frac{2N+1}{\cosh{(2r)}}\right)^{n_A}
\mathbb{E}\left[\det\left(\mathbb{I}_{n_A}+\tanh{(2r)}\mathbb{Z}_A\right)^{-1/2}\right] \nonumber \\
& \ge
\mathbb{E}\left[\det\left(\mathbb{I}_{n_A}+\tanh{(2r)}\mathbb{Z}_A\right)^{-1/2}\right] \, ,
\end{align}
where the inequality follows from the energy constraint, and it is saturated when $\cosh{2r}=2N+1$.

In the limit $N \ll 1$, we use the second-order expansion of the determinant,
\begin{equation}
\det(\mathbb{I}+\epsilon\mathbb{M}) = 1 + \epsilon\mathrm{tr}(\mathbb{M}) +
\frac{\epsilon^2}{2}\frac{\mathrm{tr}(\mathbb{M})^2-\mathrm{tr}(\mathbb{M}^2)}{2} + O(\epsilon^3) \, .
\end{equation}
By setting $\cosh{2r}=2N+1$, and noticing that $\mathrm{tr}(\mathbb{Z}_A)=0$, we get
\begin{align}
\mathbb{E}\left[\det\left(\mathbb{I}_{n_A}+\tanh{(2r)}\mathbb{Z}_A\right)^{-1/2}\right] = & \nonumber \\
& \hspace{-1.5cm} 1 + N \mathbb{E}\left[\mathrm{tr}(\mathbb{Z}_A^2)\right] + O(N^{3/2}) \nonumber \, .
\end{align}
Finally we obtain the upper bound:
\begin{equation}
\chi^\mathrm{min}_{(n,n_A)}(N) \leqslant \tilde\chi^\mathrm{min}_{(n,n_A)}(N) \leqslant 
1 + N \min \left\{ \mathbb{E}\left[\mathrm{tr}(\mathbb{Z}_A^2)\right] \right\} \, .
\end{equation}
It is worth noticing that the evaluation of the upper bound still
requires a constrained minimization, which is now independent of $N$.
In conclusion, in the region $N \ll 1$, the minimum
$\chi^\mathrm{min}_{(n,n_A)}(N)$ is bounded from above by a linear function of $N$. 
The value of the slope
\begin{equation}\label{eq:alphatilde}
\tilde\alpha_{(n,n_A)}:= \min \left\{ \mathbb{E}\left[\mathrm{tr}(\mathbb{Z}_A^2)\right] \right\}
\end{equation}
has been evaluated numerically, for several values of $n$, and is presented in Table \ref{tabN}. 
A comparison with the numerical estimation of 
\begin{equation}\label{eq:alpha}
\alpha_{(n,n_A)}:= \lim_{N\to 0} \frac{\partial\chi^\mathrm{min}_{(n,n_A)}(N)}{\partial N} \, ,
\end{equation}
also reported in Table \ref{tabN}, leads to conclude that the bound is tight.

\subsection{Saturation, $N\gg 1$}

Let us rewrite Eq.\ (\ref{class}) as
\begin{equation}\nonumber
\mathbb{V} = \frac{e^{2r}}{2} \left( \mathbb{W} + e^{-4r} \mathbb{W}' \right) \, ,
\end{equation}
where
\begin{equation}\nonumber
\mathbb{W} := \left(\begin{array}{cc}
\mathbb{X}\mathbb{X}^\mathsf{T} & -\mathbb{X}\mathbb{Y}^\mathsf{T}
\\ - \mathbb{Y}\mathbb{X}^\mathsf{T} &
\mathbb{Y}\mathbb{Y}^\mathsf{T}\end{array}\right) \, , 
\quad \mathbb{W}'
:= \left(\begin{array}{cc}
\mathbb{Y}\mathbb{Y}^\mathsf{T} & \mathbb{Y}\mathbb{X}^\mathsf{T} \\
\mathbb{X}\mathbb{Y}^\mathsf{T} & \mathbb{X}\mathbb{X}^\mathsf{T}
\end{array}\right) \, .
\end{equation}
We thus obtain
\begin{align}
\tilde\chi_{(n,n_A)} & = \frac{(N+1/2)^{|A|}}{(e^{2r}/2)^{|A|}}
\mathbb{E}\left[\det\left(\mathbb{W}_A+e^{-4r}\mathbb{W}'_A\right)^{-1/2}\right] \nonumber\\
& \hspace{-1cm} \geqslant \frac{(N+1/2)^{|A|}}{\left(2N+1-\frac{1}{2}e^{-2r}\right)^{|A|}}
\mathbb{E}\left[\det\left(\mathbb{W}_A+e^{-4r}\mathbb{W}'_A\right)^{-1/2}\right] \, , \label{eq:ineqchitilde}
\end{align}
where $\mathbb{W}_A$, $\mathbb{W}'_A$ are respectively sub-matrices of
$\mathbb{W}$, $\mathbb{W}'$. Notice that inequality (\ref{eq:ineqchitilde}) is saturated if $\cosh{2r}=2N+1$.

In the limit $N \gg 1$ (i.e., $e^{-r} \ll1$), we get
\begin{equation}
\tilde\chi_{(n,n_A)} \geqslant 2^{-n_A} \mathbb{E}\left[\det(\mathbb{W}_A)^{-1/2}\right] + O(e^{-2r}) \, .
\end{equation}
Then, by setting $\cosh{2r}=2N+1$, we obtain the upper bound
\begin{align}
\chi^\mathrm{min}_{(n,n_A)}(N) & \leqslant \tilde\chi^\mathrm{min}_{(n,n_A)} \nonumber \\ 
& \leqslant 2^{-n_A} \min \left\{ \mathbb{E}\left[\det(\mathbb{W}_A)^{-1/2}\right] \right\} \nonumber \\
& =: \beta_{(n,n_A)} \label{boundinfty} \, .
\end{align}
Notice that, also in this case, the evaluation of the right-hand side of this inequality
requires a constrained minimization, now
independent of the energy parameter $N$. 
These inequalities imply that the minimum
$\chi^\mathrm{min}_{(n,n_A)}(N)$ is bounded from above in the $N\to\infty$ limit. 
The upper bound in Eq.\ (\ref{boundinfty}) can be evaluated numerically. 
Table \ref{tabN} shows a comparison between the values of 
$\chi^\mathrm{min}_{(n,n_A)}(N)$, calculated for $N=10$ (where the saturation regime 
has been reached of all values of $n$ considered), and the numerically 
estimates of $\beta_{(n,n_A)}$.
Numerical evidence suggests that the upper bound is approached 
in the $N\to\infty$ limit. Comparison with Eq.\ (\ref{frustrb}) yields then a concrete estimate for the amount of frustration in the system.

\section{Conclusions}\label{end}

We have presented some results on entanglement frustration in
multimode Gaussian states, quantified by the minimum value of the
normalized potential of multipartite entanglement.
Entanglement frustration arises from the impossibility for a multimode Gaussian 
state of being maximally bipartite-entangled across all possible bipartitions
of the system.
It has been proven in \cite{noi} that entanglement frustration appears 
in multimode Gaussian states if $n_A \geqslant 2$, while for qubits --- for 
the case of balanced bipartion ($n_A = [n/2]$) --- it appears for $n=4$, 
$n \geqslant 8$ (the case $n=7$ is still under debate \cite{MMES1,others}).

The results obtained in this note extend the numerical analysis presented in
\cite{noi}, and put it on a more solid basis, due to the semi-analytical calculation
of the upper bounds in the low and high energy regimes.
Our numerical analysis suggests that these bounds are tight.
In particular, the calculation of a finite upper bound in the high energy regime
demonstrates that entanglement frustration remains finite in the $N\to\infty$ limit.

The results of the numerical analysis show a certain regularity in the 
estimates of the parameter $\alpha_{(n,n_A)}$.
The numerical estimates reported in Table \ref{tabN} suggest to conjecture the following relations:
\begin{align}
\alpha_{(2n_A+1,n_A)} & = n_A - 1 \, , \\
\alpha_{(2n_A,n_A)} & = \frac{2n_A}{2n_A-1} \, \alpha_{(2n_A+1,n_A)} .
\end{align}
We remark that the upper bounds are obtained by an optimization over the
matrices $\mathbb{R}$ in Eq.~(\ref{OS}), which have the property of being both symplectic 
and orthogonal, and hence define a representation of the unitary group $U(n)$
(see, e.g., \cite{ALN}).
These observations suggest that the employment of group-theoretical
methods could lead to deeper insight into the phenomenon of entanglement frustration
in multimode Gaussian states.

\section*{Acknowledgments}

The authors acknowledge stimulating discussions with G. Marmo about entanglement and the
geometry of quantum states.
The work of CL\ and SM\ is supported by EU through the FET-Open Project HIP (FP7-ICT-221899). 
PF\ and GF\ acknowledge support through the project IDEA of Universit\`a di Bari.


\begin{thebibliography}{99}

\bibitem{qmath} 
I. Bengtsson and K. Zyczkowski,
{\it Geometry of Quantum States: An Introduction to Quantum Entanglement}
(Cambridge University Press, Cambridge, 2006).

\bibitem{marmo1} 
V. I. Man'ko, G. Marmo, E. C. G. Sudarshan, F. Zaccaria, 
{\it Rep. Math. Phys.} {\bf 55} (2005), 405-422.

\bibitem{marmo2}
J. Grabowski, M. Kuoe, G. Marmo,
{\it J. Phys. A} {\bf 38} (2005), 10217-10244.

\bibitem{marmo3}
J. F. Carinena, J. Clemente-Gallardo, G.Marmo, 
{\it Theor.Math. Phys.} {\bf 152} (2007) 894-903.

\bibitem{bient} 
R. Horodecki, P. Horodecki, M. Horodecki and K. Horodecki,
{\it Rev. Mod. Phys.} {\bf 81} (2009), 865-942.

\bibitem{frustnjp} 
P. Facchi, G. Florio, U. Marzolino, G. Parisi and S. Pascazio, 
{\it New. J. Phys.} {\bf 12} (2010), 025015. 

\bibitem{CV} 
A. Ferraro, S. Olivares and M. G. A. Paris, 
{\it Gaussian states in continuous variable quantum information}
(Bibliopolis, Napoli, 2005).

\bibitem{MMES1} 
P. Facchi, G. Florio, G. Parisi and S. Pascazio,
{\it Phys. Rev. A} {\bf 77} (2008), 060304(R); 
P. Facchi, G. Florio, U. Marzolino, G. Parisi and S. Pascazio,
{\it J. Phys. A} {\bf 42} (2009), 055304. 

\bibitem{noi} 
P. Facchi, G. Florio, C. Lupo, S. Mancini and S. Pascazio,
{\it Phys. Rev. A} {\bf 80} (2009), 062311.

\bibitem{adesso} J. Zhang, G. Adesso, C. Xie and K. Peng,
{\it Phys. Rev. Lett.} {\bf 103} (2009), 070501.

\bibitem{others}
A. J. Scott, 
{\it Phys. Rev. A} {\bf 69} (2004), 052330;
A. Higuchi and A. Sudbery, 
{\it Phys. Lett. A} {\bf 273}, (2000) 213--217; 
I. D. K. Brown, S. Stepney, A. Sudbery and S. L. Braunstein, 
{\it J. Phys. A} {\bf 38} (2005), 1119;
S. Brierley and A. Higuchi, 
{\it ibid.} {\bf 40} (2007), 8455.

\bibitem{ALN}
P. Aniello, C. Lupo and M. Napolitano,
{\it Open Sys. \& Inf. Dynamics} {\bf 13} (2006), 415--426.

\end{thebibliography}
\end{document}